\documentclass[12pt]{article}
\usepackage{amssymb}
\usepackage[dvips]{epsfig}

\newcommand{\be}{\begin{equation}}
\newcommand{\ee}{\end{equation}}
\def\bea{\begin{eqnarray}}
\def\eea{\end{eqnarray}}

\newcommand{\bn}{\begin{eqnarray}}
\newcommand{\en}{\end{eqnarray}}

\newcommand{\p}{\partial}

\newcommand{\mn}{\mu\nu}

\newcommand{\nn}{\nonumber}

\newcommand{\no}{\noindent}

\newcommand{\s}{\,\,\,\,}
\def\bea{\begin{eqnarray}}
\def\eea{\end{eqnarray}}

\newcommand{\beq}{\begin{eqnarray}}
\newcommand{\eeq}{\end{eqnarray}}
\begin{document}

\title{\textbf{Spin-1 duality in $D$-dimensions}}
\author{D. Dalmazi\footnote{Partially supported by CNPq}
and R. C. Santos\footnote{Supported by CAPES} \\
\textit{{UNESP - Campus de Guaratinguet\'a - DFQ} }\\
\textit{{Av. Dr. Ariberto Pereira da Cunha, 333} }\\
\textit{{CEP 12516-410 - Guaratinguet\'a - SP - Brazil.} }\\
\textsf{dalmazi@feg.unesp.br }}
\date{\today}
\maketitle

\begin{abstract}

It is known that the Maxwell theory in $D$ dimensions can be
written in a first order form (in derivatives)  by introducing a
totally antisymmetric field which leads to a $(D-3)$-form dual
theory. Remarkably, one can replace the antisymmetric field by a
symmetric rank two tensor ($W_{\mu\nu}=W_{\nu\mu}$). Such master
action establishes the duality between the Maxwell-theory and a
fourth order higher rank model in a $D$-dimensional flat space
time. A naive generalization to the curved space shows a
connection between the recently found $D=4$ critical gravity and
the Maxwell theory plus a coupling term to the Ricci tensor
($R_{\mu\nu}A^{\mu}A^{\nu}$). The mass of the spin-1 particle
which appears in the  $D=4$ critical gravity linearized around
anti-de Sitter space is the same one obtained from the Ricci
coupling term. We also work out, in flat space time, the
explicitly massive case (Maxwell-Proca) which is dual to a second
order theory for $W_{\mu\nu}$.
\end{abstract}

\newpage

\section{Introduction}

The power of duality in field theory can hardly be underestimated,
specially if we take into account the variety of applications
 of the AdS/CFT conjecture \cite{maldacena}, see e.g. \cite{adsreview}. An
earlier example where duality has also played an important role is the rigorous proof of confinement in a four
dimensional (supersymmetric) field theory \cite{sw}. Those are examples of interacting theories.

In free (quadratic) field theories a typical approach to duality
makes use of a master action \cite{dj} which depends on two
different fields. Schematically, for massless fields, one starts
from a Lagrangian density ${\cal L}\left[A,B\right] = BB + 2\,
A\hat{D}B$ where $\hat{D}$ is some differential operator. On one
hand, the Gaussian integral over the the $B$ field furnishes
${\cal L}\left[A\right]= A \hat{D}^{\dagger}\hat{D} A$ while the
path integral over the $A$ field leads to the functional
constraint $\hat{D}B=0$ whose general solution $B(C)$ introduces
another field $C$. Back in the master action we get ${\cal
L}\left[C\right]= B(C)B(C)$ which is dual to ${\cal
L}\left[A\right]$. Along those lines the authors of \cite{op} have
shown the duality between a massless scalar field (zero-form) and
a massless $2$-form in $D=3+1$ dimensions. Later this was
generalized to $p$-form and $(D-2-p)$-form duality, see for
instance \cite{hl}.

For massive particles we have duality between a $p$-form and a
$(D-p-1)$-form. Now the master action is a bit different since we
have also a quadratic term in the $A$ field such that we can
obtain the dual theories  by Gaussian integrating either the $B$
field or the $A$ field. Usually, there is no constraint to be
solved in the massive case. We review this procedure in both
massless and massive cases in subsections 2.1 and 3.1
respectively. There we also introduce sources and determine a
local correspondence (dual map) between the dual theories which
guarantees equivalence of correlation functions up to contact
terms. Although the above dualities involve free theories, they
may suggest new interesting interacting theories as in \cite{bht}.

For our purposes it is important to consider  ${\cal
L}\left[A,B\right] $ as a lower order (in derivatives) version of
${\cal L}\left[A\right]$ and notice that there is no need of using
antisymmetric fields to decrease the order. Our starting point
here is a first order version of the Maxwell theory in $D$
dimensions obtained in \cite{kmu} with the help of a rank-two
symmetric tensor. In section 2.2 we obtain, in flat space, the
dual to the Maxwell theory by solving a functional constraint. We
compare correlation functions of gauge invariants in both dual
theories. We make some comments on a possible curved space version
of this master action. In the curved space the Maxwell action is
modified by a an interaction with the Ricci tensor:
$R_{\mn}A^{\mu}A^{\nu}$. On the dual side, a naive solution of the
constraint equation leads to a dual gravitational theory in $D=4$
which has been recently considered \cite{lupope} in the literature
and known to possess spin-1 massive particles in the spectrum
after linearization around AdS spaces \cite{bhrt}. The mass of
those spin-1 particles perfectly agrees with the one obtained in
the dual vector theory when we consider the additional term
$R_{\mn}A^{\mu}A^{\nu}$.

In section 3.2 we work out the massive formulation of the master
action of \cite{kmu} and show the duality between the
Maxwell-Proca model in a flat space with $D$ dimensions and a
second order model for the rank-two symmetric field ($S_W$). In
the last section we draw some conclusions. In the appendix we run
the Dirac-Bergmann algorithm in the Hamiltonian approach as a
double check on unitarity and the counting of degrees of freedom
for the dual $S_W$ theory.

\section{The massless case}

\subsection{$(D-3)$-form / $1$-form duality}

In this section we recall the $(D-3)$-form dual theory to the
Maxwell theory and establish a local dual map between correlation
functions in both theories.

It is known that in $D$-dimensions we can rewrite the Maxwell theory in a first-order form by using a totally
antisymmetric tensor of rank $D-2$, i.e., a $(D-2)$-form. Namely\footnote{In this work we use
$\eta_{\mu\nu}=(-,+,\cdots , +)$ and $\epsilon_{\mu_1\mu_2\cdots \mu_k \mu_{k+1} \cdots
\mu_D}\epsilon^{\mu_1\mu_2\cdots \mu_k \nu_{k+1} \cdots \nu_D} = -
 k! \, (D-k)! \det\, \delta_{\mu_i}^{\nu_j}$. Moreover for totally antisymmetric indices
 we have $\left\lbrack \alpha_1 \cdots \alpha_N\right\rbrack =
 \sum_{P} (-1)^P P\left( \alpha_1, \cdots ,\alpha_N \right)/N!$ while $(\alpha\beta) =
 \left(\alpha\beta + \beta\alpha\right)/2$ .},

\be {\cal L} \left[A,B,J\right] = -\frac{(D-2)!}4 B_{\mu_1 \cdots \mu_{D-2}} ^2 + \frac {\epsilon^{\mu_1 \cdots
\mu_D}}2 B_{\mu_1 \cdots \mu_{D-2}} \p_{\mu_{D-1}}A_{\mu_D}  + B_{\mu_1 \cdots \mu_{D-2}}J^{\mu_1 \cdots
\mu_{D-2}} . \label{lab} \ee

\no We have introduced a source term. Integrating over the
$(D-2)$-form $B_{\mu_1 \cdots \mu_{D-2}}$ in a path integral we
have the Maxwell theory plus source terms:

\be {\cal L} \left[A,J\right] = - \frac 14 F_{\mu\nu}^2 + \frac
1{2(D-2)!}\epsilon_{\mu_1 \cdots \mu_D}J^{\mu_1 \cdots
\mu_{D-2}}\p^{\mu_{D-1}}A^{\mu_D} + \frac{J_{\mu_1 \cdots
\mu_{D-2}}^2}{2(D-2)!} \label{la}\ee

\no On the other hand, if we integrate over the vector field we obtain a functional delta function enforcing the
vector constraint:

\be  \epsilon^{\mu_1 \cdots \mu_D}\p_{\mu_{D-1}}B_{\mu_1 \cdots
\mu_{D-2}} = 0 \label{c1} \ee

\no whose general solution introduces a $(D-3)$-form: $B_{\mu_1
\cdots \mu_{D-2}} = \p_{\left[\mu_{D-2}\right.}C_{\left.\mu_1
\cdots \mu_{D-3}\right]}$. Back in (\ref{lab}) we have a dual
model to the Maxwell theory:

\be  {\cal L} \left[C,J\right] = -\frac{(D-2)!}4 \left(
\p_{\left\lbrack\mu_{D-2}\right.}C_{\left.\mu_1 \cdots
\mu_{D-3}\right\rbrack}\right)^2 + J^{\mu_1 \cdots
\mu_{D-2}}\p_{\left\lbrack\mu_{D-2}\right.}C_{\left.\mu_1 \cdots
\mu_{D-3}\right\rbrack} \label{lc} \ee

\no Thus, we end up with the known duality between a $1$-form and
a $(D-3)$-form which is a particular case of the $p$-form and
$(D-p-2)$-form duality for massless particles. Furthermore, by
taking functional derivatives with respect to the source we have
the local dual map:

\be \p_{\left\lbrack\mu_{D-2}\right.}C_{\left.\mu_1 \cdots
\mu_{D-3}\right\rbrack} \leftrightarrow \frac{\epsilon_{\mu_1
\cdots \mu_{D}}\p^{\mu_{D-1}}A^{\mu_D}}{2\, (D-2)!}
\label{dualmap1} \ee

\no The map connects gauge invariant quantities in both theories ${\cal L} \left[C,J\right]$ and ${\cal L}
\left[A,J\right]$. It is such that the correlation functions of the left-hand side of (\ref{dualmap1})
calculated in the theory (\ref{lc}) agree with the correlation functions of the right-hand side of
(\ref{dualmap1}) calculated in the theory (\ref{la}) up to contact terms which have no particle content and stem
from the quadratic term in the source in (\ref{la}). The correspondence (\ref{dualmap1}) also maps the equations
of motion (in the absence of sources) of (\ref{la}) and (\ref{lc}) into each other.

\subsection{Fourth order Maxwell dual}

Remarkably, one can use instead of a $(D-2)$-form a symmetric
tensor $W_{\mn}=W_{\nu\mu}$  to rewrite the Maxwell theory in a
first-order form:

\be S\left\lbrack A,W,T \right\rbrack  = \int\, d^D\, x \left(
W^{\mn}W_{\mn} - \frac{W^2}{D-1} + 2 \, W^{\mn}\p_{(\mu}A_{\nu )}
+ W_{\mn}T^{\mu\nu}\right) \label{saw0} \ee

\no where $W=W_{\mu}^{\,\mu}$. As far as we know, the above theory
has first appeared in the appendix of \cite{kmu} in the case of
$D=4$ dimensions. Its $D$-dimensional generalization is trivial
from formulas of \cite{kmu}. We have added an external source term
$W_{\mn}T^{\mu\nu}$. Despite of depending only on the symmetric
combination $\p_{(\mu}A_{\nu )}=\left(\p_{\mu}A_{\nu} +
\p_{\nu}A_{\mu}\right)/2$, the action (\ref{saw0}) is equivalent
to the Maxwell theory. This can be made clear rewriting
(\ref{saw0}), after neglecting a surface term, as:

\bea S\left\lbrack A,W,T \right\rbrack  &=& \int\, d^D x
\left\lbrace \left\lbrack W^{\mn}  + \p^{(\mu}A^{\nu )} -
\eta^{\mn} \p \cdot A + \frac{T^{\mn}}2 - \frac{\eta^{\mn} T}2
\right\rbrack^2
\nn \right. \\
&-& \left. \frac 1{D-1}\left\lbrack W - (D-1)\left(\p \cdot A +
\frac T2\right)\right\rbrack^2 -\frac 14 F_{\mn}^2
\right. \label{saw0b} \\
 &+&  \left. T^{\mn}\left\lbrack \eta_{\mn} \p \cdot A - \p_{(\mu}A_{\nu )}\right\rbrack + \frac{T^2-T_{\mn}^2}4
\right\rbrace \nn \eea

\no After the shift $W^{\mn} = \tilde{W}^{\mn} - \p^{(\mu}A^{\nu
)}- T^{\mn}/2 + \eta^{\mn}\left( \p\cdot A + T/2 \right) $ we have
two decoupled mass terms without dynamics for the
$\tilde{W}_{\mn}$ fields plus the Maxwell theory and source terms.
After integrating over $\tilde{W}_{\mn}$ we have the Lagrangian
density:

\be {\cal L}\left\lbrack A,T \right\rbrack = -\frac 14 F_{\mn}^2 +
T^{\mn}\left\lbrack \eta_{\mn} \p \cdot A - \p_{(\mu}A_{\nu
)}\right\rbrack + \frac{T^2-T_{\mn}^2}4 \label{max} \ee

\no Formula (\ref{saw0b}) makes patent the invariance of
(\ref{saw0}), in the absence of sources,  under the gauge
transformations \cite{kmu}:

\be \delta \, A_{\mu} = \p_{\mu} \phi \quad ; \quad \delta \,
W_{\mn} = \Box \theta_{\mn}\phi \, \quad \Rightarrow \delta \, W =
(D-1)\Box\phi  \label{gt} \ee

\no Where we define the projection operators:

\be \theta_{\alpha\beta} = \left( \eta_{\alpha\beta} -
\omega_{\alpha\beta} \right) \quad , \quad \omega_{\alpha\beta} =
\frac{\p_{\alpha}\p_{\beta}}{\Box} \label{theta} \ee

On the other hand, if we start from the  action (\ref{saw0}) and integrate over the vector field  we get the
functional constraint below which plays the role of (\ref{c1}) :

\be \p^{\mu}W_{\mn} = 0 \label{c2} \ee

In $D=1+1$ the reader can check that  the general solution, linear
in fields, of (\ref{c2}), i.e.,
$W_{\mn}=\epsilon_{\mu}^{\,\,\,\delta}\epsilon_{\nu}^{\,\,\,\gamma}\p_{\delta}\p_{\gamma}
h$ is pure gauge $W_{\mn}=\Box\theta_{\mn}h$. This is in agreement
with the fact that we have $D-2$ degrees of freedom for a massless
spin-1 particle in $D$-dimensions.

In $D=2+1$, the general solution of (\ref{c2}), linear in fields,
is given by
$W_{\mn}=\epsilon_{\mu}^{\,\,\,\alpha\delta}\epsilon_{\nu}^{\,\,\,\beta\gamma}\p_{\delta}\p_{\gamma}h_{\alpha\beta}$
where $h_{\alpha\beta}=h_{\beta\alpha}$. Plugging it back in
(\ref{saw0}), at $T^{\mn}=0$, we have the linearized version of
the massless limit \cite{deserprl} of the new massive gravity
\cite{bht}, the so called K-term :

\be S^*\left[h\right] = \int\, d^3\, x \left\lbrack \sqrt{-
g}\left(R_{\mu\nu}^2 -\frac 38 R^2 \right)\right\rbrack_{hh} =
\int\, d^3\, x \, h_{\alpha\beta}\left(2 \Box
\theta^{\alpha\mu}\Box \theta^{\beta\nu} - \Box
\theta^{\alpha\beta}\Box\theta^{\mu\nu}\right)h_{\mu\nu} \label{k}
\ee

\no where $g_{\mu\nu}=\eta_{\mu\nu} + h_{\mu\nu}$. Although of fourth-order, the K-model is unitary
\cite{deserprl,unitary} and describes one massless mode in agreement with its dual theory (Maxwell) which is on
its turn equivalent to a massless scalar theory in $D=2+1$. The duality between the K-term and the Maxwell
theory in $D=2+1$ is not new \cite{more}. However, in \cite{more} the K-model appears upon integration over a
symmetric rank-2 tensor while the Maxwell action is obtained via solution of a constraint equation, in this
sense the master action (1) is dual to the corresponding one of \cite{more}. We also notice that the gauge
symmetry $\delta W_{\mn} = \Box \theta_{\mn}\phi$ follows from the linearized Weyl symmetry $\delta h_{\mn} =
\eta_{\mn} \phi$ of (\ref{k}).

For arbitrary dimensions $D \ge 2$ the general solution of
(\ref{c2}), linear in fields, is given by

\be W_{\mn}=\epsilon_{\mu}^{\,\,\,\alpha_1 \cdots
\alpha_{D-2}\delta}\epsilon_{\nu}^{\,\,\,\beta_1 \cdots
\beta_{D-2}\gamma}\p_{\delta}\p_{\gamma}h_{\left\lbrack
\alpha_1\cdots \alpha_{D-2}\right\rbrack \left\lbrack\beta_1
\cdots \beta_{D-2}\right\rbrack} \label{gs} \ee

\no where $h_{{\left\lbrack\alpha_1\cdots
\alpha_{D-2}\right\rbrack \left\lbrack\beta_1 \cdots
\beta_{D-2}\right\rbrack}} = h_{\left\lbrack\beta_1\cdots
\beta_{D-2}\right\rbrack \left\lbrack\alpha_1 \cdots
\alpha_{D-2}\right\rbrack}$.

Alternatively, we can write, as in the $D=3+1$ case treated in
\cite{dst}, the general solution (\ref{gs}) as

\be W_{\mn}= \p^{\delta}\p^{\gamma}B_{\mu\delta\nu\gamma} \label{gs2} \ee

\no where

\be B_{\mu\delta\nu\gamma} =
\epsilon_{\mu\delta}^{\s\,\,\,\alpha_1 \cdots \alpha_{D-2}}
\epsilon_{\nu\gamma}^{\s\,\,\,\beta_1 \cdots
\beta_{D-2}}h_{\left\lbrack \alpha_1\cdots
\alpha_{D-2}\right\rbrack \left\lbrack\beta_1 \cdots
\beta_{D-2}\right\rbrack} \label{bfield} \ee

\no Substituting the general solution in (\ref{lab}) we have a
dual fourth-order description of the Maxwell theory in any
dimension $D \ge 3$:

\be S^*\left[ h \right] = \int d^D \, x \left\lbrack
\left(\p^{\delta}\p^{\gamma}B_{\mu\delta\nu\gamma}\right)^2 -
\frac{\left(\p^{\delta}\p^{\gamma}B^{\mu}_{\,\,\,\delta\mu\gamma}\right)^2}{D-1}
+ \p^{\delta}\p^{\gamma}B_{\mu\delta\nu\gamma} T^{\mu\nu}
\right\rbrack \label{sh} \ee

\no where $B_{\mu\delta\nu\gamma}(h)$ is given in (\ref{bfield}).
In $D=3+1$ the same theory was obtained before \cite{dst} in a
different approach and shown to be unitary by means of a canonical
analysis. In $D=2+1$, as we have already mentioned, the theory is
also unitary. Based on the master action (\ref{saw0}) we believe
that unitarity will hold in arbitrary dimensions as a consequence
of the unitarity  of the Maxwell theory and the fact that
(\ref{gs}) is the general solution of (\ref{c2}).

As we have done in the last subsection, see (\ref{dualmap1}), we
can compare correlation functions of gauge invariants in both
theories, i.e., Maxwell, see (\ref{max}), and its dual $S^*\left[
h \right] $.  Perhaps, the most natural invariant under (\ref{gt})
in the dual theory $ S^*\left[ h \right] $ is $\p^{\mu}W_{\mn}$.
However, due to the functional constraint (\ref{c2}) its
correlation functions are trivial (vanish). Later we will see that
they correspond to correlation functions of
$-\p^{\mu}F_{\mu\nu}(A)/2$ in the Maxwell theory up to contact
terms. In searching for a nontrivial gauge invariant, we consider
that on the Maxwell side the basic local gauge invariant is the
antisymmetric tensor ($F_{\alpha\beta}$) and on the dual side we
must take at least one derivative of $W_{\mn}$ in order to have
invariance under (\ref{gt}). So we can think of taking linear
combinations of $\p_{\alpha}W_{\beta\nu}- \p_{\beta}W_{\alpha\nu}
$ and $ \left(\eta_{\alpha\nu} \p_{\beta}-\eta_{\beta\nu}
\p_{\alpha}\right) W$. Along this way we end up with the gauge
invariant below which has nonvanishing correlation functions in
the dual theory $ S^*\left[ h \right] $

\be G_{[\alpha\beta ]\nu} \equiv \p_{\alpha}W_{\beta\nu}-
\p_{\beta}W_{\alpha\nu} + \frac{\left(\eta_{\alpha\nu}
\p_{\beta}-\eta_{\beta\nu} \p_{\alpha}\right)}{D-1} W \quad . \ee

\no Replacing the source term $T^{\mu\nu}W_{\mn}$ by a new one
$T^{[\alpha \beta]\nu}G_{[\alpha\beta ]\nu}$ in (\ref{saw0}), it
is clear that we can keep the gauge invariance of the action
without requiring any constraint on the sources. Integrating by
parts one derivative we can rewrite the source term once again in
the form $\tilde{T}^{\mu\nu}W_{\mn}$ where $\tilde{T}^{\mu\nu}$
contain combinations of one derivative of $T^{[\alpha \beta]\nu}$,
so we can still use the action (\ref{max}) replacing $T^{\mu\nu}$
by $\tilde{T}^{\mu\nu}$. This procedure leads to the dual map,

\be G_{[\alpha\beta ]\nu} \leftrightarrow - \frac 12
\p_{\nu}F_{\alpha\beta}  \quad , \label{dualmap2} \ee

\no  Thus, correlation functions of $G_{[\alpha\beta ]\nu} $ in
the dual theory $S^*\left[ h \right] $ correspond to correlation
functions of $\p_{\nu}F_{\alpha\beta}/2$ in the Maxwell theory up
to contact terms which are due to quadratic terms in $T^{\mu\nu}$
in (\ref{max}). From those correlation functions we can infer the
correlations of $F_{\mn}$.

Now we finish this section commenting on a possible curved space
generalization of the $S^*\left[ h \right]$/Maxwell duality. The
natural curved space version of the master action (\ref{saw0}), in
the absence of sources, is given by:

\bea S\left\lbrack A,W,T \right\rbrack  &=& \int\, d^D \, x
\sqrt{-g} \left( W^{\mn}W_{\mn} - \frac{W^2}{D-1} + 2 \,
W^{\mn}\nabla_{(\mu}A_{\nu )} \right) \nn\\ &=& \int \, d^Dx
\sqrt{-g} \left\lbrace \left( W^{\mn} + \nabla^{(\mu }A^{\nu )} -
g^{\mu\nu} \nabla \cdot A \right)^2 - \frac{\left\lbrack W - (D-1)
\nabla \cdot A \right\rbrack^2}{D-1} \right. \nn
 \\  &-& \left. \frac 14 F_{\mu\nu}^2 + R_{\mn}A^{\mu}A^{\nu}
\right\rbrace \nn \\
&=& \int\, d^D \, x \sqrt{-g} \left(
\tilde{W}^{\mn}\tilde{W}_{\mn} - \frac{\tilde{W}^2}{D-1}  -\frac
14 F_{\mu\nu}^2 + R_{\mn}A^{\mu}A^{\nu} \right) \label{ricci} \eea

\no Where $ \nabla_{\mu}$ is the curved space covariant derivative
and $\tilde{W}_{\mn} = W_{\mn} + \nabla_{( \mu}A_{\nu )} -
g_{\mu\nu} \nabla \cdot A $.

The Ricci tensor  $R_{\mu\nu}$ has appeared due to the
non-commutativity of the covariant derivatives:
$\nabla_{\mu}\nabla_{\nu}A^{\nu} = \nabla_{\nu}\nabla_{\mu}A^{\nu}
- R_{\mu\nu}A^{\nu}$. In (\ref{ricci}) we have two trivial
(non-dynamic) terms for $\tilde{W}_{\mn}$ decoupled from the
vector field. After neglecting those trivial terms we end up with
the Maxwell theory plus a Ricci ``mass term'' in the curved space.
So, the minimal coupling to gravity in the master action
(\ref{saw0}) originates a non-minimal coupling for the massless
vector field which breaks the $U(1)$ gauge symmetry.

On the other hand, integrating over the vector field $A_{\mu}$ in
the first line of (\ref{ricci}) we have the curved space version
of the  constraint (\ref{c1}):

\be \nabla^{\mu}W_{\mn} = 0 \label{c3}\ee

 \no We do
not know its general solution but we can certainly begin with the
lowest order terms in derivatives of the metric $W_{\mn}(g) = a_0
\, g_{\mn} + a_2 \left( R_{\mn} - g_{\mn}R/2 \right) + a_4
K_{\mu\nu} + \cdots $ where $a_0,a_2,a_4,\cdots $ are arbitrary
constant coefficients. The tensor $K_{\mu\nu}=(1/\sqrt{-g})\delta
S_4/\delta g^{\mu\nu}$ is obtained from a general fourth-order
Lagrangian density which can be written as ${\cal L}_4 = \alpha
R^2 + \beta R_{\mu\nu}^2 + \gamma {\cal L}_{GB}$, where
$\alpha,\beta,\gamma$ are arbitrary constants and the Gauss-Bonnet
term is ${\cal L}_{GB}= R_{\mu\nu\alpha\beta}^2 -4 R_{\mu\nu}^2 +
R^2$. Plugging the solution $W_{\mn}(g)$ back in the first line of
(\ref{ricci}) we get:

\be S\left\lbrack W_{\mn}(g) \right\rbrack = \int d^D x \,
\sqrt{-g} \left\lbrace \frac {a_0^2 D}{D-1} + a_0a_2
\frac{D-2}{D-1} R + a_2^2 \left\lbrack R_{\mn}^2 -
\frac{D}{4(D-1)}R^2 \right\rbrack - \frac{2 a_0 a_4}{(D-1)} K +
\cdots \right\rbrace \label{critical} \ee

\no where the dots stand for terms of sixth or higher order in
derivatives of the metric and

\be K = g^{\mu\nu}K_{\mu\nu} = \frac{4-D}2 \left( \alpha R^2 +
\beta R_{\mu\nu}^2 + \gamma {\cal L}_{GB} \right) + \left\lbrack
2\alpha (D-1) + \frac{\beta\, D}2 + 4 \gamma \right\rbrack
\nabla_{\mu}\nabla^{\mu} R  - 4 \gamma
\nabla_{\mu}\nabla_{\nu}R^{\mn} \label{K} \ee

\no The last two terms of (\ref{K}) are total derivatives which
can be neglected in (\ref{critical}) for arbitrary $D$-dimensions.
In $D=4$ we can discard $K$ completely and (\ref{critical})
becomes exactly, dropping the dots, the critical gravity theory
recently found in \cite{lupope}.

Upon linearization around an AdS background $R_{\mn}=\Lambda g_{\mn}$ the authors of \cite{bhrt} have shown the
existence of ``Proca-log-modes'' in the critical gravity theory in $D=4$. Those spin-1 massive modes can be
derived from a curved space Lagrangian density, in the notation\footnote{The notation of \cite{bhrt} can be
recovered by matching the coefficients of the terms proportional to $a_0 a_2$ and $a_2^2$. This leads  to
$a_2=1/(\kappa m\sqrt{D-2})$ and $a_0=m\sigma (D-1)/(\kappa \sqrt{D-2})$, up to an overall sign. The coefficient
of the $a_o^2$ term (cosmological term) in (\ref{critical}) comes out correctly without any fit just like the
relative coefficient between $R_{\mn}^2$ and $R^2$ inside the term proportional to $a_2^2$.} of \cite{bhrt}, of
the Maxwell-Proca form ${\cal L}_{MP}= -F_{\mn}^2/4 + 3 m^2 \sigma\, A^{\mu}A_{\mu}$. This is in full agreement
with (\ref{ricci}) since the criticality condition \cite{lupope} requires $\Lambda = 3\sigma m^2$ in the
notation of \cite{bhrt}. A unitary theory requires the unusual sign $\sigma < 0$ for the Einstein-Hilbert term
as in the $D=3$ case \cite{more} which is an earlier example of a critical gravity.

Regarding the general case of critical gravity in $D$-dimensions
(see \cite{dllpst}) we must mention that the mass of the Proca
modes predicted by (\ref{ricci}) is still in agreement with the
results obtained for the critical gravity but in the absence of
the Gauss-Bonnet term where the criticality condition becomes, in
the notation of \cite{bhrt}, $\Lambda = (D-1)\sigma m^2$. From the
point of view of (\ref{critical}) we must set $a_4=a_6=\cdots =0$.
Apparently, the key point is to make sure that upon linearization
the solution $W_{\mn}(g)$ is in fact a general solution to the
constraint (\ref{c3}) without redundancies. We believe that
possible redundancies can be eliminated by field redefinitions.

%

We finish this section by  mentioning that, alternatively, in
order to keep the $U(1)$ gauge invariance in the curved space we
could have added the non-minimal coupling term with negative sign
$-R_{\mu\nu}A^{\mu}A^{\nu}$ to the first line of (\ref{ricci})
such that we end up with the pure Maxwell theory in the curved
space after the shifts in the $W_{\mn}$ fields. At the level of
master action the $U(1)$ gauge invariance would be restored in the
curved space.  However, due to this new term there would be no
functional constraint equation for the $W_{\mn}$ fields and we
could in principle Gaussian integrate over the vector field and
obtain a dual theory containing  the exotic term
$\nabla^{\mu}W_{\mn}(R^{-1})^{\nu\beta}\nabla^{\alpha}
W_{\alpha\beta}$ which involves the inverse of the Ricci tensor.
It is not yet clear if this a consistent solution, even for
special backgrounds, to the gravitational coupling problem
mentioned in \cite{dst}. We are still investigating this
possibility.

\section{The massive case}

\subsection{$(D-2)$-form/$1$-form duality}

This subsection parallels the subsection 1.1. Adding a mass term
and a source term for the vector field in (\ref{lab}) we have the
master action, see \cite{botta},

\bea {\cal L}_m (A,B) &=& -\frac{(D-2)!}4 B_{\mu_1 \cdots
\mu_{D-2}} ^2 + \frac 12 \epsilon^{\mu_1 \cdots
\mu_D}B_{\mu_1 \cdots \mu_{D-2}} \p_{\mu_{D-1}}A_{\mu_D}  \nn\\
&+& B_{\mu_1 \cdots \mu_{D-2}}J^{\mu_1 \cdots \mu_{D-2}} - \frac
{m^2}2 A_{\mu}A^{\mu} + J_{\mu}A^{\mu} \label{labm} \eea

\no Integrating over $B_{\mu_1 \cdots \mu_{D-2}}$ in the path
integral we have the Maxwell-Proca theory plus source dependent
terms:

\bea {\cal L}_m (A) &=& -\frac 14 F{\rm}^2 - \frac {m^2}2 A_{\mu}A^{\mu} + J_{\mu}A^{\mu} \nn\\
&+& \frac{\epsilon^{\mu_1 \cdots \mu_D} J_{\mu_1 \cdots \mu_{D-2}}\p_{\mu_{D-1}}A_{\mu_D}}{(D-2)!} + \frac
{J^2_{\mu_1 \cdots \mu_{D-2}}}{(D-2)!} \label{lam} \eea

\no On the other hand, integrating over the vector field in (\ref{labm}) we have the Kalb-Ramond $(D-2)$-form
dual model:

\bea {\cal L}_m (B) &=& - \frac{\left\lbrack
(D-1)!\right\rbrack^3}{4\, m^2} \left( \p_{\left\lbrack \mu_1
\right.}B_{\left.\mu_2 \cdots \mu_{D-1}\right\rbrack }\right)^2
-\frac{(D-2)!}4
B_{\mu_1 \cdots \mu_{D-2}}^2 \nn\\
&+& B_{\mu_1 \cdots \mu_{D-2}}J^{\mu_1 \cdots \mu_{D-2}} -
\frac{\epsilon^{\mu_1 \cdots \mu_D}J_{\mu_D}\p_{\mu_{D-1}}B_{\mu_1
\cdots \mu_{D-2}}}{m^2} + \frac{J_{\mu}J^{\mu}}{m^2} \label{lbm}
\eea

\no Comparing (\ref{lam}) with (\ref{lbm})  we can calculate
correlation functions in the Kalb-Ramond model in terms of
correlations in the Maxwell-Proca theory and vice-versa by using
the dual maps below respectively:

\be B_{\mu_1 \cdots \mu_{D-2}} \leftrightarrow
\frac{\epsilon_{\mu_1 \cdots
\mu_{D-2}}^{\quad\quad\quad\mu_{D-1}\mu_D}\p_{\mu_{D-1}}A_{\mu_D}}{(D-2)!}
\label{dualmap3} \ee

\be A^{\mu}  \leftrightarrow -\frac{\epsilon^{\mu_1 \cdots \mu_{D-1}\mu}\p_{\mu_{D-1}}B_{\mu_1 \cdots
\mu_{D-2}}}{m^2} \label{dualmap4} \ee

\no Notice that in the massive case we do not need to worry about gauge invariance. The maps (\ref{dualmap3})
and (\ref{dualmap4}) are consistent with each other up to contact terms as expected. This completes the
$1$-form/$(D-2)$-form duality which is a special case of the $p$-form/$(D-1-p)$-form duality for massive
theories.

\subsection{$S_W$/Maxwell-Proca duality}

Similarly we can define the massive version of (\ref{saw0}),

\be S_m \left\lbrack W,A \right\rbrack  = \int\, d^D\, x \left(
W^{\mn}W_{\mn} - \frac{W^2}{D-1} + 2 \, W^{\mn}\p_{(\mu}A_{\nu )}
- \frac{m^2}2 A^{\mu}A_{\mu} + J \cdot A \right) \label{s1} \ee

\no which can be written as

\bea S_m &=& \int \, d^Dx  \left\lbrace \left( W^{\mn} + \p^{(\mu }A^{\nu )} - \eta^{\mu\nu} \p \cdot A
\right)^2 - \frac{\left\lbrack W - (D-1) \p \cdot A \right\rbrack^2}{D-1} \right. \nn
 \\  &-& \left. \frac 14 F_{\mu\nu}^2 - m^2 \frac{A^2}2 + J \cdot A
\right\rbrace \label{proca} \eea

\no Thus, after a shift in $W_{\mn}$,  we have the Proca action
plus some decoupled mass terms for $W_{\mn}$ which can be dropped,
whereas integrating over the vector field and rescaling $W_{\mn}
\to m \, W_{\mn}/\sqrt{2}$ we get

\be S_m = S_W + \int \, d^D \, x  \left\lbrack \frac{J^2}{2 m^2} -
\frac{\sqrt{2}\, J^{\mu}\p^{\nu} W_{\mn}}{m^2} - \frac{m^2}2
\left(A_{\mu} + \frac{2\p^{\nu} W_{\mn}}{m} -
\frac{J_{\mu}}{m^2}\right)^2\right\rbrack \label{sm2} \ee

\no where

\be S_W = \int \, d^D \, x \left\lbrack (\p^{\nu}W_{\mn})^2 + \frac{m^2}2 \left( W_{\mn}^2 - \frac{W^2}{D-1}
\right) \right\rbrack \label{sw} \ee

\no After a shift in $A_{\mu}$ in (\ref{sm2}) it is clear that
$S_W$ must be dual to the Maxwell-Proca theory. Indeed, from
(\ref{sw}) we can read off the propagator :

\bea \left\langle W^{\lambda\mu}(x) W_{\alpha\beta}(y) \right\rangle &= & \left\lbrace\frac{P_{SS}^{(2)}}{m^2} -
\frac{P_{SS}^{(1)}}{\Box - m^2} + \left\lbrack \frac{2\Box \, (D-1)}{m^2} + 2 - D \right\rbrack
\frac{P_{SS}^{(0)}}{m^2} \right. \nn\\
&-& \left. \frac{\sqrt{D-1}}{m^2} \left(P_{SW}^{(0)} +
P_{WS}^{(0)} \right)\right\rbrace^{\lambda\mu}_{\alpha\beta} \quad
, \label{propa} \eea

\no where the differential operators are defined as \be \left( P_{SS}^{(2)}
\right)^{\lambda\mu}_{\s\s\alpha\beta} = \frac 12 \left( \theta_{\s\alpha}^{\lambda}\theta^{\mu}_{\s\beta} +
\theta_{\s\alpha}^{\mu}\theta^{\lambda}_{\s\beta} \right) - \frac{\theta^{\lambda\mu} \theta_{\alpha\beta}}{D-1}
\quad , \label{ps2} \ee

\be \left( P_{SS}^{(1)} \right)^{\lambda\mu}_{\s\s\alpha\beta} = \frac 12 \left(
\theta_{\s\alpha}^{\lambda}\,\omega^{\mu}_{\s\beta} + \theta_{\s\alpha}^{\mu}\,\omega^{\lambda}_{\s\beta} +
\theta_{\s\beta}^{\lambda}\,\omega^{\mu}_{\s\alpha} + \theta_{\s\beta}^{\mu}\,\omega^{\lambda}_{\s\alpha}
 \right) \quad , \label{ps1} \ee

\be \left( P_{SS}^{(0)} \right)^{\lambda\mu}_{\s\s\alpha\beta} = \frac 1{D-1} \,
\theta^{\lambda\mu}\theta_{\alpha\beta} \quad , \quad \left( P_{WW}^{(0)} \right)^{\lambda\mu}_{\s\s\alpha\beta}
= \omega^{\lambda\mu}\omega_{\alpha\beta} \quad , \label{psspww} \ee

\be \left( P_{SW}^{(0)} \right)^{\lambda\mu}_{\s\s\alpha\beta} = \frac 1{\sqrt{D-1}}\,
\theta^{\lambda\mu}\omega_{\alpha\beta} \quad , \quad  \left( P_{WS}^{(0)}
\right)^{\lambda\mu}_{\s\s\alpha\beta} = \frac 1{\sqrt{D-1}}\, \omega^{\lambda\mu}\theta_{\alpha\beta} \quad ,
\label{pswpws} \ee

\no It turns out that single and double poles of (\ref{propa}) at
$\Box =0$ cancel out exactly such that $\left\langle W^{\mn}(x)
W_{\alpha\beta}(y)\right\rangle $ is analytic at $\Box =0$ and
contains only one single pole at $\Box = m^2$ in the spin-1
sector. In order to check the particle content of $S_W$ we look at
the saturated two point amplitude

\bea A_W(k) &=& T^*_{\mu\nu}(k) \left\langle W^{\mn}(-k)W_{\gamma\beta} (k) \right\rangle
T^{\gamma\beta}(k) \nn\\
&=& - \frac{2\, i}{k^2(k^2
+m^2)}\left(k^{\epsilon}T_{\epsilon\mu}^* \theta^{\mu\nu}
k^{\lambda}T_{\lambda\nu} + \cdots \right) \label{aw} \eea

\no where $T_{\mu\nu}(k)$ are symmetric sources which are analytic functions of the momentum. The dots stand for
analytic functions at $k^2 = - m^2$. It is instructive to compare (\ref{aw}) with the corresponding amplitude of
the Maxwell-Proca theory:

\bea A_{{\rm MP}}(k) &=& J^^*_{\mu}(k) \left\langle A^{\mu}(-k)A^{\nu} (k) \right\rangle
J_{\nu}(k) \nn\\
&=& i \left(\frac{J^*_{\mu}(k)\theta^{\mu\nu}J_{\nu}(k)}{k^2 +
m^2} + \frac{J^*_{\mu}(k)\omega^{\mu\nu}J_{\nu}}{m^2}\right)
\label{aproca} \eea

\no As in (\ref{aw}) the pole at $k^2=0$ cancels out in (\ref{aproca}). The imaginary part of the residue at
$k^2=-m^2$ is positive as expected:

 \be R_m = \Im \lim_{k^2 \to - m^2} \left(k^2 + m^2\right) A_{{\rm
 Proca }}(k) = J_t^* \cdot J_t \label{rm} \ee

 \no where $J_t$ is the transverse part of the source, $k \cdot
 J_t=0$. Choosing $k_{\mu} = (m,0,0)$ it becomes clear that $J_t^* \cdot
 J_t = \vert J_t^1 \vert^2 + \vert J_t^2 \vert^2 > 0 $.
 Analogously, identifying $ k^{\lambda}T_{\lambda\mu}
 \leftrightarrow J_{\mu}$ it is easy to see that $R_m = \Im \lim_{k^2 \to - m^2} \left(k^2 + m^2\right) A_w(k) > 0
 $. Therefore, both $S_W$ and $S_{{\rm Proca}}$ have exactly the
 same particle content, one massive spin-1 mode in a
 $D$-dimensional space-time ($D\ge 2$). In $D=3+1$ the action $S_W$ has appeared before in
 \cite{pvn} as a special case of  a general expression for a second-order
 action quadratic in symmetric rank-2 tensors and restricted to be
 unitary. Here we are showing that $S_W$ is dual to to the Proca
 theory for any $D$. The linear terms in the source $J_{\mu}$, after the trivial shifts in $W_{\mn}$ and $A_{\mu}$  in
 (\ref{proca}) and (\ref{sm2}), reveal the dual map

 \be A_{\mu} \leftrightarrow - \frac{\sqrt{2}}m \p^{\nu} W_{\nu\mu} \label{dualmap5}
 \ee

 \no which allows us to compute correlation functions in the
 Maxwell-Proca theory from the $S_W$ action. Conversely, the correlations of the
 fundamental field $W_{\mn}$ can be obtained from the Maxwell-Proca
 theory via the map

 \be W_{\mn} \leftrightarrow  \frac{\sqrt{2}}m \left\lbrack \eta_{\mu\nu} \p \cdot A -
 \p_{(\mu}A_{\nu )} \right\rbrack \label{dualmap6} \ee

\no which could be derived by introducing sources
 for $W_{\mn}$, see (\ref{max}). Those maps are similar to (\ref{dualmap3}) and (\ref{dualmap4}).
 As before, the
 correlation functions in the dual theories must match
 up to contact terms. We can give a simple argument to show the consistency of (\ref{dualmap5}) and
 (\ref{dualmap6}) as follows.
 Substituting the left hand side of (\ref{dualmap6})
 in (\ref{dualmap5}) we get the Maxwell-Proca equation
 of motion (\ref{procaeq}). From an integral of a total (functional) derivative

\be \int {\cal D}A \frac{\delta}{\delta A(x_1)} \left\lbrack
e^{i\, S} A(x_2) \cdots A(x_N)\right\rbrack =0 \label{ward} \ee

\no where symbolically $S=\int d^D x A \hat{{\cal O}} A/2 $, we can derive $\left\langle \hat{{\cal O}} A
(x_1)A(x_2) \cdots A(x_N)\right\rangle = 0 $ whenever $x_1 \ne x_j $ for all $j=2,\cdots, N$. So the equation of
motion $\hat{{\cal O}} A = 0 $ is enforced in the correlation functions if we neglect coinciding points (contact
terms), which is exactly when the dual maps are supposed to hold.

Regarding the classical equivalence between the Maxwell-Proca
theory  and the $S_W$ model, from the equations of motion:

 \be \frac{\delta S_W}{\delta W^{\mn}} =
 \p_{\nu}\p^{\alpha}W_{\mu\alpha} +
 \p_{\mu}\p^{\alpha}W_{\nu\alpha} + m^2 \left( \eta_{\mu\nu}
 \frac{W}{D-1} - W_{\mn} \right)=0 \label{eqm} \ee

\no we can derive $\p^{\mu}\p^{\nu}W_{\mn} = -m^2 W/(2(D-1)) $ and

\be \Box\left( \p^{\alpha}W_{\nu\alpha} \right) -
\p_{\nu}\left(\p^{\alpha}\p^{\beta}W_{\alpha\beta}\right) - m^2
\p^{\alpha}W_{\alpha\nu} =0 \label{procaeq1}\ee

\no which is equivalent to the Maxwell-Proca equation with the
identification (\ref{dualmap5}):

\be \Box \, A_{\nu} - \p_{\nu} \left( \p \cdot A \right)  - m^2
A_{\nu} = 0\label{procaeq} \ee

\no From (\ref{procaeq}) we can derive the transverse condition $\p \cdot A =0 $  and the Klein-Gordon equation
$(\Box - m^2)A_{\mu} = 0 $ which describe a spin-1 massive particle. Since (\ref{procaeq1}) has been derived
from (\ref{eqm}) by applying a derivative one might wonder whether the general solution of (\ref{eqm}) contains
more information than the transverse condition and the Klein-Gordon equation. In order to answer that question
we start with a general Ansatz for a symmetric rank-2 tensor: $W_{\mu\nu} = \p_{\mu} f_{\nu}  + \p_{\nu} f_{\mu}
+ W_{\mu\nu}^{(T)} $ where $W_{\mu\nu}^{(T)} $ is given in (\ref{gs}) and $f_{\mu}$ is arbitrary. From
(\ref{procaeq1}), which follows from (\ref{eqm}), we deduce $\p^{\mu}\p^{\nu}W_{\mn} = 0$, consequently due to
$\p^{\mu}\p^{\nu}W_{\mn} = -m^2 W/(2(D-1)) $, which also follows from (\ref{eqm}), we have $W=0$. So, back in
(\ref{eqm}) we conclude that $W_{\mu\nu}$ is purely longitudinal, i.e., $W_{\mu\nu}^{(T)} = 0 $. Substituting
$W_{\mu\nu} = \p_{\mu} f_{\nu}  + \p_{\nu} f_{\mu}$ in (\ref{eqm}) we obtain the third order equation

\be \left(\Box - m^2 \right) \p_{(\mu}f_{\nu )} + \p_{\mu}\p_{\nu} \left(\p \cdot f \right) = 0 \label{eqm2} \ee

\no From $W=0$  we deduce $\p \cdot f = 0$ and from (\ref{eqm2}) we have $\p_{\mu} g_{\nu} + \p_{\nu} g_{\mu} =
0 $ where $ g_{\mu} \equiv \left(\Box - m^2 \right) f_{\mu} $. Assuming that the fields vanish at infinity, the
solution is $g_{\mu} =0 $. Consequently, we conclude that (\ref{eqm}) describes a massive spin-1 particle and
nothing else in agreement with our pole analysis of (\ref{propa}). In the appendix we perform the Dirac-Bergman
constraints analysis for $S_W$. We end up with  $D-1$ degrees of freedom and a positive definite Hamiltonian as
required for a unitary massive spin-1 particle.

Since we have rescaled $W_{\mn} \to \, \sqrt{2} W_{\mn}/m $ it is worth looking at the massless limit of $S_W$.
At $m=0$ we have the local gauge invariance $\delta_{\Lambda} W^{\mn} = \epsilon^{\mu\alpha_1 \cdots
\alpha_{D-2}\delta}\epsilon^{\nu\beta_1 \cdots \beta_{D-2}\gamma}\p_{\delta}\p_{\gamma}\Lambda_{\left\lbrack
\alpha_1\cdots \alpha_{D-2}\right\rbrack \left\lbrack\beta_1 \cdots \beta_{D-2}\right\rbrack}$. By adding a
symmetry breaking term  with an arbitrary coefficient $\lambda $ we have

\be {\cal L}_W^{m=0} = \left(\p^{\nu}W_{\mn}\right)^2 + \frac{\lambda}2 \left(\p^{\mu} W\right)^2 \label{lw} \ee

\no which allows us to obtain

\be \left\langle W^{\mn}(x) W_{\alpha\beta}(y)\right\rangle_{m=0}
 = - \left\lbrace \frac{2\, P_{SS}^{(1)}}{\Box} + \frac
{\left(P_{SW}^{(0)} + P_{WS}^{(0)} \right)}{\Box\sqrt{D-1}}  -
\frac{P_{WW}^{(0)}}{\Box } - \left(2 +
\frac{\lambda}{D-1}\right)\frac { P_{SS}^{(0)}}{\lambda\Box}
\right\rbrace^{\mu\nu}_{\alpha\beta} \label{propa2} \ee

\no It contains single and double poles at $\Box = 0$. However, if
we add a source term and require gauge invariance
$\delta_{\Lambda} \left( W_{\mn}T^{\mu\nu} \right) =0 $, the
source (in momentum space) must be of the form $T^{\mu\nu} (k) =
k^{\mu} J^{\nu}(k) + k^{\nu} J^{\mu}(k) $ where $J_{\mu}$ does not
need to be conserved. The saturated (gauge invariant) two-point
amplitude gets contribution only from the $P_{SS}^{(1)}$ and $
P_{WW}^{(0)} $ pieces (which are $\lambda$-independent). At the
end the poles cancel out and we are left with an analytic
function:

\be A^{m=0}_W(k) = \frac i2 T_{\mu\nu}^*(k) \left\langle
W^{\mn}(-k) W_{\alpha\beta}(k)\right\rangle_{S_W(m=0)}
T^{\alpha\beta}(k)
 = -2\, i \, J_{\mu}^*(k) J^{\mu} (k) \quad . \label{aw0} \ee

\no Thus, the $m=0$ limit of $S_W$ has no particle content in any
dimension $D$. This is  in agreement with the study of \cite{pvn}
for the case $D=3+1$. In the Hamiltonian formalism one can show
that $S_W$ at $m=0$ has enough first class constraints to gauge
away all degrees of freedom.

\section{Conclusion}

We have reviewed the usual duality between a $1$-form  and a $(D-3)$-form (massless case) and between a $1$-form
and a $(D-2)$-form (massive case) in subsections 2.1 and 3.1 respectively. We have established a local dual map
in both cases, see (\ref{dualmap1}) and the pair (\ref{dualmap3}) and (\ref{dualmap4}) respectively. They allow
us to calculate correlation functions of local physical  quantities in one theory in terms of the dual
quantities in the dual theory, up to contact terms.

In subsection 2.2, starting from a master action recently obtained in \cite{kmu}, we have derived a fourth order
(in derivatives) dual model to the Maxwell theory in arbitrary $D$-dimensions. In particular, in $D=2+1$ the
corresponding dual theory is the massless limit, see \cite{deserprl}, of the linearized new massive gravity of
\cite{bht}. Once again we have determined a dual map between gauge invariants in both theories.

Remarkably, a naive curved space version of the master action of \cite{kmu} leads, on one hand to the Maxwell
action plus an interacting term with the Ricci tensor ($R_{\mu\nu}A^{\mu}A^{\nu}$). On the other hand, the
``dual theory'' (\ref{critical}) (neglecting higher than fourth order terms) corresponds in $D=4$ to a critical
gravity theory which was recently found in \cite{lupope}. It contains curvature square terms with fine tuned
coefficients plus a fine-tuned cosmological term and the usual Einstein-Hilbert action. The linearized theory
around an AdS background contains \cite{bhrt} Proca modes (spin-1) whose mass matches exactly the one obtained
here via master action. However, a remark is in order. Namely, the linearized AdS critical gravity of
\cite{lupope} contains also spin-2 modes \cite{bhrt}. It is not clear how our naive generalization of the master
action of \cite{kmu} to curved spaces could be improved in order to encompass the spin-2 modes appropriately on
both sides of the duality.

In subsection 3.2 we have generalized the flat space master action
of \cite{kmu} by adding an explicit mass term for the vector field
(Proca term). Integrating over the vector field we have obtained a
second order dual theory to the Maxwell-Proca action in terms of a
symmetric rank-2 tensor. We have found the dual map between those
models in arbitrary $D$-dimensional flat space-time with $D \ge 2$
and checked that the dual theory indeed shares the same particle
content of the Maxwell-Proca theory by analyzing the analytic
structure of the symmetric tensor propagator. In the appendix we
confirm that one has $(D-1)$ degrees of freedom in the Hamiltonian
approach by running the Dirac-Bergmann algorithm. We also show
that the Hamitonian is definite positive. The massless limit of
this higher-rank description of spin-1 particles has no particle
content.

\section{Acknowledgements}

We thank Elias L. Mendon\c ca for useful discussions and bringing \cite{bhrt} to our knowledge.

\section{Appendix}

Here, we analyze the Hamiltonian constraints generated by the
massive action (\ref{sw}) , dual to the Maxwell-Proca model in
$D$-dimensions.

Since $W_{\mn}=W_{\nu\mu}$, in order to avoid unnecessary constraints we work only with independent phase space
variables $W_{\mn}$ and $\pi^{\mu\nu}$ with $\mu \le \nu$. From the Lagrangian density,

\bea \mathcal{L} &=&
\partial^{\nu}W_{\mu \nu}\partial_{\alpha} W^{\mu\alpha}
+ \frac{m^2}{2}(W_{\mu \nu}W^{\mu \nu} - \frac{W^2}{D-1}) \nn\\
&=& -(-\partial_{0}W_{00} +
\partial_{i} W_{0i})^2 + (-\partial_{0}W_{0i} +
\partial_{j} W_{ij})^2 + \frac{m^2}{2}(W_{\mu
\nu}W^{\mu \nu} - \frac{W^2}{D-1}), \label{lap1} \eea

\no we can calculate the conjugated momenta:

\begin{equation}
\pi^{00} = \frac{\partial\mathcal{L}}{\partial(\partial_0 W_{00})}
\approx 2(-\partial_{0}W_{00} + \partial^{i} W_{0i}) \qquad
\Longrightarrow \qquad
\partial_{0}W_{00} = -\frac{\pi^{00}}{2} +
\partial^iW_{0i},
\end{equation}
\begin{equation}
\pi^{0i} = \frac{\partial\mathcal{L}}{\partial(\partial_0 W_{0i})}
\approx -2(-\partial_{0}W_{0i} + \partial^{j} W_{ij}) \qquad
\Longrightarrow \qquad
\partial_{0}W_{0i} = \frac{\pi^{0i}}{2} +
\partial^jW_{ij},
\end{equation}
\begin{equation}
\pi^{ij} = \frac{\partial\mathcal{L}}{\partial(\partial_0 W_{ij})}
\approx 0   \quad ; \quad i \le j  \label{cij} \end{equation}

\no In (\ref{cij}) we have  $\frac{D(D-1)}{2} $ primary
constraints. Now, we can calculate the canonical Hamiltonian:

\begin{equation}
H_c = \int d^{D-1}x [\pi^{00}\partial_0 W_{00} +
\pi^{0i}\partial_0W_{0i} + \pi^{ij}\partial_0 W_{ij} -
\mathcal{L}] ,
\end{equation}
\begin{equation}
H_c = \int d^{D-1}x [-\frac{(\pi^{00})^2}{4} + \pi^{00}\partial^i W_{0i} + \frac{(\pi^{0i})^2}{4} +
\pi^{0i}\partial^jW_{ij} - \frac{m^2}{2}(W_{\mu \nu}W^{\mu \nu} - \frac{W^2}{D-1})]
\end{equation}
The total Hamiltonian is:
\begin{equation}
H_t = H_c + \int d^{D-1}x \lambda_{ij} \pi^{ij}.
\end{equation}
Where the sums above exist only for $i \le j$. Using the canonical
Poisson brackets for the $D(D-1)$ phase space variables:
 \begin{equation}
\{W_{\mu \nu}(x),W^{\rho \lambda}(y)\}=0; \qquad \{\pi_{\mu
\nu}(x),\pi^{\rho \lambda}(y)\}=0; \qquad \{W_{\mu
\nu}(x),\pi^{\rho \lambda}(y)\}=
\delta_{\mu}^{\rho}\delta_{\nu}^{\lambda}\delta (x-y)
\end{equation}

\no We have the secondary constraints:
 \begin{equation}
\chi^{lk} \equiv \dot{\pi}^{lk} = \partial^k \pi^{0l} + \partial^l
\pi^{0k} + 2m^2W_{lk} \approx 0 \quad ; \quad l<k \quad ,
\label{clk}
\end{equation}

 \begin{equation}
\chi^{k}\equiv\dot{\pi}^{kk} = \partial^k \pi^{0k} + m^2(W_{kk} -
\frac{W}{D-1}) \approx 0 \quad ; \quad k=1,2, \cdots D-1
\label{ckk}
\end{equation}

\no In (\ref{ckk}) there is no sum over $k$. The consistency
conditions $\dot{\chi}^{lk} = \left\lbrace \chi^{lk} , H_t
\right\rbrace \approx 0  $ fix the coefficients $\lambda_{lk} \, ,
\, l < k$ and do not generate new constraints. while
$\dot{\chi}^{k}=0$ determines all $\lambda_{kk}$ (no sum) except
the sum $\lambda_{jj}$. This can be easily seen from the
combination

\be \chi \equiv \sum_{k=1}^{D-1} \chi^{k} = \p^j\pi^{0j} + m^2
W_{00} \approx 0  \label{chi} \ee

\no The consistency equation $\dot{\chi}\approx 0$ leads to
another (tertiary) constraint:

\be \dot{\chi}\approx   \left( \nabla^2 - \frac{m^2}2\right)
\pi^{00} - m^2 \p^i W_{0i} \equiv \phi \approx 0 \label{phi} \ee

\no where $\nabla^2 = \p_j\p^j $. Finally, rewriting $\dot{\phi}
\approx 0 $ with the help of (\ref{ckk}) and (\ref{chi}) we have:

\be  W = - W_{00} + W_{jj} \approx 0 \label{psi} \ee

\no From $W\approx 0$  the constraint $\chi^{k} \approx 0 $ can be
written as (no sum)

\be  \p^k \pi^{0k} + m^2 W_{kk} \approx 0 \quad , \quad k=1,2,
\cdots D-1 \label{ckk2} \ee

In summary, adding up
(\ref{cij}),(\ref{clk}),(\ref{ckk}),(\ref{phi}) and (\ref{ckk2})
we have $D(D-1)+2$ independent second class constraints. Thus, we
end up with $2(D-1)$ unconstrained phase space variables which
correspond to $(D-1)$ degrees of freedom as expected for a massive
spin-1 particle in $D$-dimensions. In particular, we can eliminate
all variables in term of $(W_{0k},\pi^{0j})$. Back in the total
Hamiltonian, after some cancelations we have:

\bea H_t &=& \int d^{D-1}x \left\lbrack m^2\frac{\p^j W_{0j}\left(
\nabla^2 - 3 m^2/4 \right)\p^j W_{0j}}{\left(\nabla^2
- \frac{m^2}2\right)^2} + m^2 W_{0j}W_{0j} \right. \nn \\
&+& \left. \frac{\left(\pi^{0j}\right)^2}4 +
\frac{\sum_{j<k}\left(\p^j\pi^{0k}- \p^k\pi^{0j}\right)^2}{4\,
m^2} \right\rbrack \quad . \label{ht2} \eea

\no Decomposing $W_{0j} = W_j^T + W_j^L$, where

\be W_j^T = \theta_{jk}W_{0k} \quad ; \quad W_j^L =
\omega_{jk}W_{0k} \quad , \label{tl} \ee

\no we can rewrite $H_t$ in a explicitly definite positive form as expected from our proof of unitarity

\bea H_t &=& \int d^{D-1}x \left\lbrack \frac{m^4}4
\frac{W_j^L\left(m^2 - \nabla^2\right)W_j^L}{\left(\nabla^2
- \frac{m^2}2\right)^2} + m^2 W_j^TW_j^T \right. \nn \\
&+& \left. \frac{\left(\pi^{0j}\right)^2}4 +
\frac{\sum_{j<k}\left(\p^j\pi^{0k}- \p^k\pi^{0j}\right)^2}{4\,
m^2} \right\rbrack \quad . \label{ht3} \eea

\no The positiveness of the first term in (\ref{ht3}) can be
checked by integrating by parts the factor $\nabla^2$ in the
numerator or by going to the momentum space (Fourier transform).
Notice that the poles at $\nabla^2 =0$ present in the projection
operators $\theta_{ij}$ and $\omega_{ij}$ cancel out in
(\ref{ht3}) in agreement with (\ref{ht2}).

We have checked that, by using the appropriate Dirac brackets, the
reduced phase space Hamiltonian (\ref{ht3}) leads indeed to the
equations of motion (\ref{eqm}).

\end{document}